\documentclass{iopart}
\usepackage{graphicx}
\begin{document}
\jl{6}
\title[No statistical excess]{No statistical excess in EXPLORER/NAUTILUS observations in the year 2001}
\author{Lee Samuel Finn\footnote{Also Center for Gravitational Physics and
    Geometry, Department of Physics and Department of Astronomy and
    Astrophysics; e-mail LSFinn@PSU.Edu}} 
\address{Center for Gravitational Wave Physics, The Pennsylvania State
  University, 104 Davey Laboratory, University Park, PA 16802} 
\pacs{0480, 0430}
\begin{abstract}
A recent report on gravitational wave detector data from the NAUTILUS
and EXPLORER detector groups claims a statistically significant excess
of coincident events when the detectors are oriented in a way that
maximizes their sensitivity to gravitational wave sources in the
galactic plane. While not claiming a detection of gravitational waves,
they do strongly suggest that the origin of the excess is of
gravitational wave origin. In this note we show that the statistical
analysis that led them to the conclusion that there is a statistical
excess is flawed and that the reported observation is entirely
consistent with the normal Poisson statistics of the reported detector
background.

\end{abstract}

\maketitle
\section{Introduction}
The NAUTILUS and EXPLORER gravitational wave detector group has
recently reported an examination of 90 days of data taken in 2001
\cite{astone02a}. From each detector's data they identify events,
which they calibrate in terms of excitation energy. Over the 90 days
of observation they identify 31 events that are coincident between the
two detectors, both in terms of arrival time and excitation
energy. They bin these 31 events in terms of the sidereal hour of
event arrival and, separately, in terms of the solar hour of the event
arrival. The expected number of events in each bin owing to background
is Poisson. The number of background events in each bin is estimated
by examining events that are coincident following an unphysical time
lag. Examining the sidereal and solar hour binning Astone et
al. \cite{astone02a} claim
\begin{itemize}
\item there is a statistically significant excess of events
  corresponding to sidereal hour 3;  
\item an even more significant excess when the sidereal hour 3 and 4
  bins are combined; and  
\item the event distribution with respect to sidereal hour is
  significantly different than the binning with respect to solar hour.  
\end{itemize}
Since the detectors have a non-trivial detector pattern, an event
excess at certain sidereal hours, but none when the data are binned by
solar hour, would be suggestive of detector excitations associated
with gravitational radiation incident from a fixed direction in the
sky. The authors draw attention to this, stating that they ``wait for
a stronger confirmation of our
result before reaching any definite
conclusion and claiming that gravitational waves have
been observed''
\cite[next to final paragraph]{astone02a}. Here we show that
\begin{itemize}
\item the number of observed events arriving in sidereal hour 3 is not
  statistically in excess of the number expected from the Poisson
  statistics of the reported detector background; 
\item the number of observed events in the combined hour 3 and 4 pair
  are not statistically in excess of the number expected from the
  Poisson statistics of the reported detector background and the beam
  (i.e., the antenna pattern) of the two detectors; and 
\item 
the event distribution in solar hour and sidereal are statistically
indistinguishable.
\end{itemize}
We conclude that \emph{the published observations do not support the
published conclusion} that the observations are inconsistent with
Poisson statistics, or that there is a statistical difference between
the distribution of events in sidereal hour vs. solar hour, or that
there is an event excess associated with times when the detectors are
favorably oriented with respect to the galactic plane.

\section{The observations and their interpretation}

\subsection{Sidereal hour 3}

Table \ref{tbl:obs} summarizes the observations that the
EXPLORER/NAUTILUS detector group cite as demonstrating a statistically
significant excess in sidereal hour 3 \cite[figure
5]{astone02a}.\footnote{The tabulated numbers were taken from an
electronic version of the figure, which was enlarged and superposed
with a grid.} In sidereal hour 3 they observe four events when their
background estimate suggests an average of 0.9 events. (No estimate of
the uncertainty in the background is provided in \cite{astone02a} and,
while a matter of concern for that analysis, we will accept the
implication that the expected number of background events in each hour
is known to a high precision.) They go on to note that when the
expected number of events in a Poisson process is 0.9 the probability
of observing four or more events is quite small:
\begin{equation}
P(n>3|0.9) = 1-e^{-0.9}\sum_{n=0}^3\frac{(0.9)^n}{n!} = 1.35\%
\end{equation}
On this point their claim of a statistical excess in the number of
events hinges.

Neglected, however, is the fact that they have undertaken 24 separate
experiments, corresponding to the 24 different sidereal hours. The
correct question to ask is whether, over these 24 different
experiments, there is a signficant probability that at at least one
hour will be exceptional at this level: i.e., that at least one of
these 24 Poisson experiments will have a number of events in excess of
the expected number at the 1.4\% level.  This probability is
\begin{equation}
p = 1-\left[1-P(n>3|0.9)\right]^{24} = 27.8\%;
\end{equation}
i.e., if we were to simply draw from a Poisson distribution 24 times
(or from 24 different Poisson processes once), we would expect an
``excess'' at this level slightly more than 1 in 4 times. On this
basis the observations reported in \cite{astone02a} do not support the
conclusion that there is unexpectedly large number of events in
sidereal hour 3, or in any sidereal hour bin, contrary to the claim in
\cite[\S6]{astone02a}.

Context is critical in assessing the significance of a data feature
identified through analysis. The more independent tests one performs
on a random data set, the more likely it is that something
``apparently'' significant will turn-up: i.e., \emph{every} random
data will show ``unusual'' characteristics if looked at in enough
different ways. Mistakenly identifying such ``unusual
characteristics'' as significant is sometimes referred to as the
cluster illusion. In this case we see that the ``cluster'' of events
at sidereal hour 3 is an illusion: noise alone would produce this
outcome more frequently than one in four observations. More
colloquially, the cluster illusion can be thought of as shooting the
arrow first and drawing the bullÕs eye later.

\subsection{Sidereal hours 3 and 4}

A second claim made in \cite{astone02a} is that the combination of
hour 3 with hour 4 (corresponding to 7 events when 1.7 are expected)
is significant, though hour 4 is not, by itself, noteworthy. This
claim of significance is not made quantitative in \cite{astone02a},
through they emphasize its potential significance by the \emph{a
posteriori} reasoning that the detectors are at these times more
sensitive to radiation originating in the galactic plane (though not
the galactic center) than at other times.

As previously discussed, the significance of this, or any other,
observation about the data cannot be judged independently of its
context. Astone et al. \cite{astone02a} did not set-out to examine
adjacent two-hour combinations, nor did they bin the data in two hour
intervals. Instead, having  found what they considered
to be an unusual hour they subsequently noted that an adjacent hour 
had an expected 
count of 0.8
and an actual count of 3 and claimed this as giving added 
significance to the claim
of an event excess at or about sidereal hour 3.

It is next to impossible to evaluate the significance of a hypothesis
like this one, which is framed \emph{a posteriori}: i.e., after the
data have been examined and contingent on what was found in that first
examination. To do so requires that we enumerate all the alternative
hypotheses that would have been suggested by all possible
observations: an impossible task. For this reason the results of ``testing''
\emph{a posteriori}
hypotheses are generally considered inadmissible as evidence for or 
against any claim. 

Nevertheless, it is still instructive to consider the hour 3/4 pairing
and attempt to bound its significance. That starting point is the
relevance of the \emph{pairing} of hours 3 and 4.  That pairing was
suggested \emph{after and because} hour 3 was (mistakenly) identified
as significant. The probability of flipping a coin and obtaining heads
is not made more or less likely if, on the previous throw of the coin,
heads was also obtained; similarly, because the pairing was noted both
\emph{after and because} hour 3 was identified as ``significant'',
\emph{the significance of the count in the hour 4 bin is not enhanced
by its being adjacent to hour 3.} An alternative way of expressing the
same point is that knowing that hour 3 had four counts one shouldn't
be surprised that the combined count of hour 3 and hour 4 is greater
still.  Correspondingly, \emph{hour 4 stands alone.}

A proper evaluation of the significance of the hour 4 count requires
an understanding of all of the equivalent observations about the data
that, had they occurred, would have been considered
significant. Exactly because the hour 4 count was identified \emph{a
posteriori} this is difficult to do. For guidance, we turn to the
\emph{a posteriori} justification given for judging the hour 3/4
pairing significant. Counting all hours similar to hour 4 under this
reasoning will give us a lower-bound on the number of equivalent
hypotheses and, thus, an upper-bound on the significance of the
particular hour 4 count.

\begin{figure}
\begin{center}
\includegraphics[height=10cm]{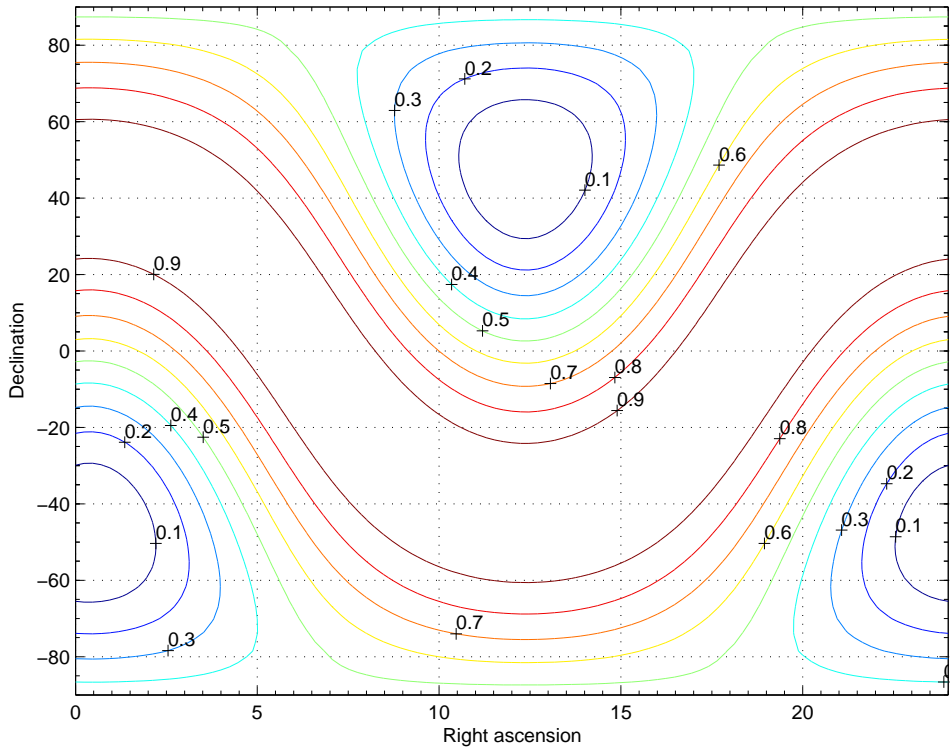}
\end{center}
\caption{Contours of mean amplitude sensitivity to points in the sky
  at sidereal hour 3. Note that the detector beam is quite broad;
  correspondingly, a distribution of sources in the galactic plane can
  be expected to yield an elevated count in any or all of several
  hours surrounding the hour of peak sensitivity.}\label{fig:sense} 
\end{figure}

\begin{figure}
\begin{center}
\includegraphics[height=10cm]{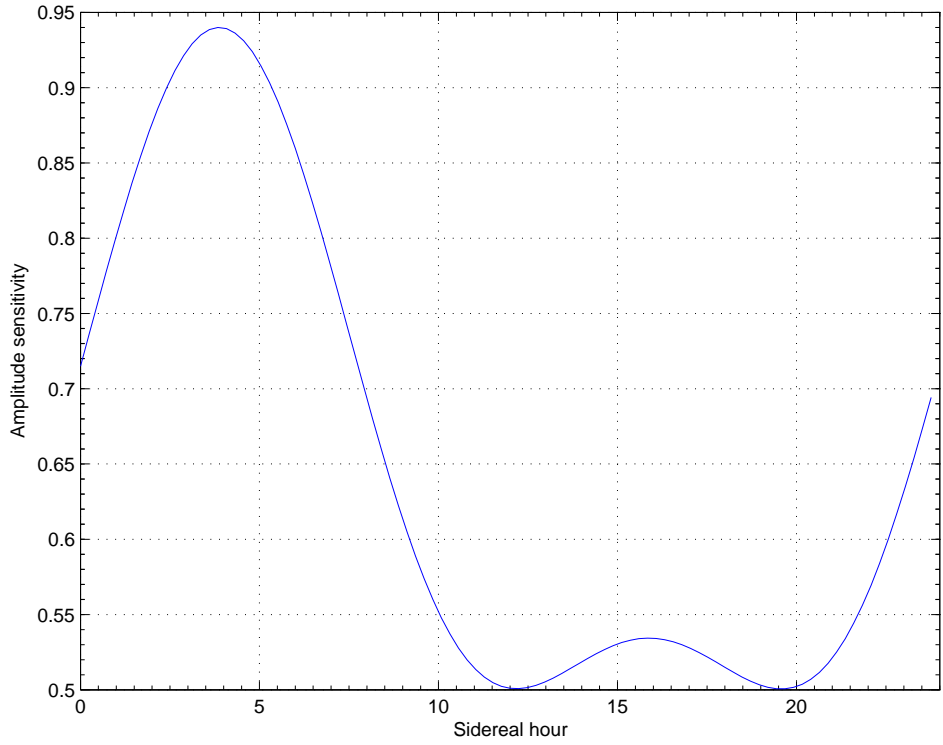}
\end{center}
\caption{Overlap of the mean detector beam with the galactic plane as
  a function of detector sidereal hour. For sources distributed
  throughout the galactic plane the amplitude sensitivity in sidereal
  hours 1--6 will not vary by more than 10\%.}\label{fig:overlap} 
\end{figure}

The \emph{a posteriori} rationale for focusing on the hour 3/4
combination is its relative sensitivity to galactic plane sources.
Important to this point is that bar detectors like EXPLORER and
NAUTILUS have rather broad beams (i.e., antenna patterns). In
particular, owing just to the antenna beam (i.e., ignoring the angular
width of the galactic plane) an offset of $\pm2$ hours of right
ascension from ``optimal'' overlap with the galactic plane reduces the
amplitude sensitivity to sources randomly distributed throughout the
galactic plane by only 5\%. Indeed, over the entire 24 hour sidereal
hour period the sensitivity of the detector pair to equidistant
sources in the galactic plane never falls below 50\%: see figures
\ref{fig:sense} and \ref{fig:overlap}.

Since the focus is on galactic plane sources we assume that the
hypothesis to be tested involves sources at distances sufficient that
they are in fact distributed in a two-dimensional disk (as opposed to
so near that the distribution is isotropic). Assuming that the bursts
are standard candles, half of the sources observed with strain
amplitude above the detector amplitude threshold $h_0$ will have a
strain amplitude at the detector greater than $2^{1/3}h_0$ and 75\% of
the sources will have an amplitude greater than $1.1h_0$. As is
apparent from figure \ref{fig:overlap} the amplitude sensitivity of
the detector varies by no more than 10\% over hours 1--6;
consequently, 75\% of the events observable when the detector is at
peak sensitivity (approximately sidereal hour 3.5) would also have
been above threshold in any of hours 1--6.

Thus the \emph{a posteriori} explanation for the combined count in the
hour 3 and 4 bins would have fit equally well a combination of hour 3
and any of hours 1, 2, 4, 5 or 6. In this set of five hours the
background can be expected to yield an hour like hour 4 with
probability
\begin{equation}
p = 1-\left[1-P(n>2|0.8)\right]^5 = 21.6\%;
\end{equation}
i.e., one in five times. Correspondingly, hour 4 is not exceptional by
any of the usual standards of evidence for a significant result.

It is important to recognize that this is an upper limit on the
significance of the hour 4 count: the actual significance is likely
lower because we have not taken into account other associations,
identified \emph{a posteriori}, that might have been justified through
other \emph{a posteriori} reasoning. Based on this minimum number of
possible associations we have evaluated the \emph{maximum}
significance of the noted association. This is sufficient to show that
there is no statistical excess associated with favorable detector
orientation relative to the galactic plane.

\subsection{Comparison of sidereal and solar hour distributions}

A third claim made in \cite{astone02a} is that the distribution of
events when binned according to sidereal hour is significantly
different then when binned according to solar hour. This claim, too,
is not made quantitative in \cite{astone02a}; however, we can examine
it here. Given the Poisson parameter describing the detector
background and the number of observed events that they report we can
calculate the probability of the particular observation of number of
events in the 24 different sidereal (solar) hours that they have made,
assuming Poisson statistics with the background rate they have
estimated:
\begin{equation}
P\left(\{n_k: k = 0\ldots23\}|\{\lambda_k:k=0\ldots23\}\right) =
\prod_{k=0}^{23} P(n_k|\lambda_k)
\end{equation}
where
\begin{equation}
P(n|\lambda) = \frac{\lambda^n}{n!}e^{-\lambda}
\end{equation}
is the Poisson distribution, $n_k$ is the observed number of events in
sidereal (solar) hour $k$ and $\lambda_k$ is the expected number of
events in sidereal (solar) hour k.  Focus on the ratio of the
probability of the particular observation reported on in
\cite{astone02a} when binned by sidereal hour to the probability when
binned by solar hour:
\begin{eqnarray}
p &=& \frac{P_{\rm{sid}}}{P_{\rm{sol}}}\\
P_{\rm{sid}} &=& 
P\left(\{n_{\rm{sid},k}: k =
0\ldots23\}|\{\lambda_{\rm{sid},k}:k=0\ldots23\}\right)\\ 
P_{\rm{sol}} &=&
P\left(\{n_{\rm{sol},k}: k =
0\ldots23\}|\{\lambda_{\rm{sol},k}:k=0\ldots23\}\right). 
\end{eqnarray}
For the particular observation reported in \cite{astone02a} and
summarized in table \ref{tbl:obs} the observed value of $p$ is 20.5\%:
i.e., assuming nothing more than Poisson statistics at the background
rate, binning the reported observations in sidereal hours yields a
distribution about five times less likely than binning in solar hours.

\begin{figure}
\begin{center}
\includegraphics[height=10cm]{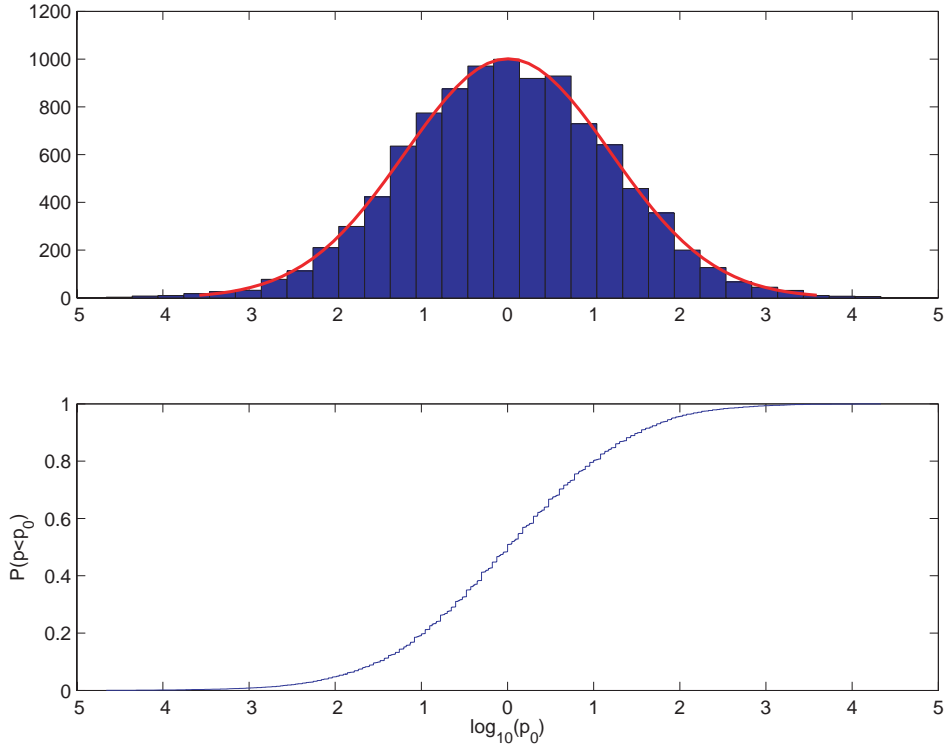}
\end{center}
\caption{As a measure of the difference between a sidereal hour and
  solar hour binning of the EXPLORER/NAUTILUS data reported in
  \protect\cite{astone02a} we can calculate the ratio of the
  probability of observing each binning, assuming only Poisson
  statistics, and ask how frequently we would expect to determine a
  smaller value. This figure shows the result of a Monte Carlo
  simulation that distributes 31 events, equal to the number observed
  in the EXPLORER/NAUTILUS coincidence run, randomly in time and then
  calculates the ratio of the probabilities associated with
  corresponding sidereal and solar hour binning. Ten thousands trials
  are represented in this Monte Carlo. The top panel shows the
  histogram in $\log_{10}$ of the probability ratio, with a line
  indicating a Gaussian fit. The bottom panel shows the probability
  that the observed probability ratio would be less than the given
  value. The observed value of the probability ratio (20.5\%, or
  $\log_{10}(20.5\%)=-0.69$) is entirely consistent with the noise
  associated with the distribution.}\label{fig:mc} 
\end{figure}

To determine the significance of this result we can perform a Monte
Carlo simulation, distributing 31 events (the total number observed
and reported on in \cite[figure 5]{astone02a}) randomly in time,
binning the results by sidereal hour and by solar hour, calculating
$p$ for this simulated observation of a Poisson process and computing
the distribution of $p$ that emerges. Figure \ref{fig:mc} shows the
results of such a Monte Carlo made under the simplifying assumption
that the Poisson parameter is the same for all bins and equal to 31/24
(corresponding to 31 events distributed over 24 hours). The first
panel of Figure \ref{fig:mc} shows a histogram of the distribution of
$\log_{10}p$ and the second panel shows the probability that $p$ is
less than some $p_0$. From these two figures it is clear that a value
of $p$ less than 20.5\% is expected approximately 28\% of the time:
i.e., more frequently than one in every four experiments when the
observations are entirely of detector background noise. The claim that
the there is a difference in the distribution of events when binned by
solar or sidereal hour is thus unsupported by the data.

Before leaving this claim behind it is worth noting that the
comparison of sidereal and solar hour distributions is not entirely
independent of the identification of an excess in a particular
sidereal hour, as is apparent from the definition of the odds ratio in
terms of the probabilities associated with the observed number in each
hour bin.

\section{Discussion}
The interpretation of features identified in a data set cannot be
separated from the context in which they were identified. As the
number of independent hypotheses increases the likelihood of a large
fluctuation leading to a false alarm in at least one also increases;
correspondingly, the threshold for a significant result must also be
increased. Astone \cite{astone02a} sought first to identify whether
the observed counts in each of 24 sidereal hours was consistent with
the Poisson statistics of the background. They overestimated the
significance of the observed count in sidereal hour 3 by separating it
from the context of the 23 other independent hypotheses they tested on
the EXPLORER/NAUTILUS coincidence data. By treating the count in each
hour as if it were the only hypothesis being tested on the data they
underestimated their actual false alarm rate.

The context necessary for determining the significance of a data set
feature is unambiguous for \emph{a priori} hypotheses: i.e.,
hypotheses clearly articulated in advance of the analysis. On the
other hand, it is next to impossible to determine the necessary
context for testing \emph{a posteriori} hypotheses: i.e., hypotheses
that arise after the analysis is underway and are suggested by the
outcome of \emph{a priori} hypotheses. A proper accounting of all the
possible \emph{a posteriori} hypotheses that might have been suggested
by equivalent \emph{post hoc} reasoning is required to put the
significance of an \emph{a posteriori} hypothesis on a proper
footing. Attempts to do so almost always fall prey to the so-called
``Texas Sharpshooter Fallacy'', so-named after the proverbial Texas
``sharpshooter'' who would fire first and draw the bull's eye later.

It is for this reason that it is common to draw the distinction
between confirmatory and exploratory data analysis. Confirmatory
analysis involves only \emph{a priori} hypotheses with significance
levels set with full understanding of the context of the
analysis. Their goal is to test the \emph{a priori} hypotheses. The
results of properly executed confirmatory analyses are statistically
sound and reliable. Exploratory analyses, on the other hand, are
fishing expeditions. Their proper goal is not to test hypotheses, but
to suggest new hypotheses to be tested in later confirmatory
analyses. Since one can't test hypotheses identified in exploratory
analyses in the data set in which that are identified conclusions
regarding them are rarely published without confirmatory analysis. The
most common exception to this rule are those rare circumstances when
something truly unexpected stands-out so clearly that it remains
strongly ``significant'' by any reasonable attempt to account for the
context of the hypothesis. That is not the case with any of the
hypotheses identified by Astone et al.

\section{Conclusions}

The observations reported in \cite{astone02a} are entirely consistent
with the estimates of the detector noise background: no result
reported rises beyond the level of a ``1 $\sigma$'' event. The
mistaken claim of an excess in sidereal hour 3 arises from a failure
to consider the number of trials involved in binning the data into 24
one hour bins. When a proper account of the trials is taken into
account, the observation of the number of events in hour 3 is seen to
be entirely consistent with the reported detector background by all
usual standards of evidence. The claim that the hour 3/4 combination
is also large, and of which much is made through \emph{a posteriori}
reasoning, is seen to be poorly framed and, on closer examination, to
also be consistent with the reported detector background. Finally,
when the binning of the observations into sidereal hours and solar
hours are compared the particular distribution of events is also seen
to be consistent with reported statistics of the detector
background. We conclude that there is no statistical evidence to
support the suggestions of an event excess in the reported NAUTILUS
and EXPLORER data.

\ack

I gratefully acknowledge discussions with Nils Andersson, Albert
Lazzarini, Peter Saulson and Phil Wilhelms. This work was supported by
National Science Foundation awards PHY~01-14375 and PHY~00-99559.

\begin{table}
\caption{Background rates and observed number of events corresponding
  to sidereal and solar hours from figure 5 of
  \cite{astone02a}.}\label{tbl:obs} 
\begin{indented}
\item[]\begin{tabular}{@{}l|llll}
\br
&\multicolumn{2}{c}{Sidereal}&\multicolumn{2}{c}{Solar}\\
Hour&Poisson&Observed&Poisson&Observed\\
&param.&number&param.&number\\
\mr
0& 0.6&2&1.00&2\\
1&1.0&1&1.30&2\\
2&1.1&0&1.20&1\\
3& 0.9&4&0.90&0\\
4& 0.8&3&1.00&0\\
5&0.7&1&1.30&2\\
6& 0.6&1&1.20&2\\
7&0.6&1&1.00&1\\
8&0.7&0&0.90&1\\
9&1.0&1&0.50&1\\
10& 0.5&0&0.60&2\\
11& 0.7&1&0.80&1\\
12& 1.5&1&0.60&0\\
13& 1.6&2&1.30&2\\
14& 2.2&3&1.50&1\\
15& 0.9&0&1.50&2\\
16& 1.0&1&1.00&1\\
17& 1.3&1&0.80&1\\
18& 1.3&2&1.10&1\\
19& 0.9&1&1.10&3\\
20&1.7&2&0.80&1\\
21& 1.5&1&0.90&0\\
22& 1.1&1&0.90&2\\
23& 1.1&1&1.20&2\\
\end{tabular}
\end{indented}
\end{table}

\end{document}